\documentstyle[aps,psfig]{revtex}
\begin{document}
\draft
\preprint{MKPH-T-02-01}
\title{
\hfill{\small {\bf MKPH-T-02-01}}\\
{\bf The $\eta$-$3N$ problem with separable interactions}
\footnote{Supported by the Deutsche Forschungsgemeinschaft (SFB 443)}}
\author{A.\,Fix\,$^a$ and H.\,Arenh\"ovel\,$^b$}
\address{$^a$Tomsk Polytechnic University, 634004 Tomsk, Russia}
\address{$^b$Institute f\"ur Kernphysik,
Johannes Gutenberg-Universit\"at Mainz, D-55099 Mainz, Germany}
\date{\today}
\maketitle

\begin{abstract}
The $\eta$-$3N$-interaction is studied within the four-body scattering
theory adopting purely separable forms for the two- and three-body 
subamplitudes, limiting the basic two-body interactions to $s$-waves only. 
The corresponding separable approximation for the integral
kernels is obtained by using the Hilbert-Schmidt procedure. Results are
presented for the $\eta$-$^3$H scattering amplitude and for the total elastic
cross section for energies below the triton break-up threshold.
\end{abstract}

\pacs{PACS numbers:  13.60.Le, 21.45.+v, 25.20.Lj}


\section{Introduction}
\label{intr}
In the last ten years the 
interaction of $\eta$-mesons with few-nucleon systems has attracted 
considerable interest, and quite a large amount of research has been 
carried out both on the experimental as well as on the theoretical 
side. The major aim of this activity is to obtain a
well-defined conceptual picture of the low-energy $\eta$-nuclear
interaction which then may serve to clarify the foundation of the more
fundamental $\eta N$ problem.

Of special interest is the $\eta$-$3N$-system, for which exactly soluble 
models are available. At the same time it encloses a much wider range of 
phenomena than the more simple $\eta NN$
case. In this context we would also like to mention several precise 
experimental
investigations of $\eta$-production on three-body nuclei performed
during the last decade \cite{Peng89,May96,Kru02}. 
Previous theoretical studies in this area were 
restricted to various approximations in treating the few-particle
aspects of this problem. As an earlier nonperturbative
calculation, we would like to refer to the optical model approach
of~\cite{Wilkin93}, whose purpose was to estimate qualitatively the
possible formation of bound $\eta$-meson states with the lightest nuclei. 
More refined calculations were 
presented in~\cite{Wyce95}, where the authors were able to sum the
multiple scattering series for the $\eta$-$^3$H amplitude,
including several important corrections to the trivial optical
limit. Another result reported in~\cite{Belya95} were obtained within
the so-called finite-rank-approximation (FRA) which has recently been
applied also to the
$\eta$-production from three-body nuclei~\cite{Shev01,Khem02}. 
The crucial point of this model is the neglect of target
excitations during the interaction with the 
$\eta$-meson. Clearly this assumption allows one to avoid the complications
associated with the direct solution of the four-body dynamical
equations. Concerning the present study, we were guided by the
idea that the approximations used in many-body physics, such as the
optical model or adiabatic treatment of the target, may fail when
few-particle systems are studied. Especially, this seems to be true for small
kinetic energies, where the unitarity conditions of the scattering
matrix, nucleon recoil and other effects become significant. Their
neglect, tolerated in various many-body approaches, may affect 
drastically the quality
of the few-body results. Therefore, the present work is intended
to solve the $\eta$-$3N$-problem without making any such not well
controlled approximations. 

The basis of our calculation is the four-body formalism in momentum space.
Although a variety of methods for solving the n-particle problem has been 
proposed in the literature, the Faddeev-Yakubovsky theory~\cite{Yak67,Fad70}
and the one of Alt, Grassberger and Sandhas (AGS)~\cite{GS67,AGS70} are 
most convenient and preferred for practical applications. Adopting the 
separable 
representation for the driving two-body potentials as well as for the 
subamplitudes appearing in the (1+3)- and (2+2)-partitions of the 
four-body system, both approaches lead to the same set of effective two-body
equations~\cite{GS67,Fons86,Narod81}. As is well known,  
the separable approximation of the integral kernels permits one to 
represent the dynamical equations in terms of particle exchange
diagrams. Due to this 
tractability and its relatively simple numerical realization, this
method has received wide acceptance in few-body physics. Thus at
present, a feasible formalism of four-particle theory has been
extensively developed, despite the much more complex structure of the
corresponding equations compared to the three-body case. After the 
work of Tjon~\cite{Tjon75}
impressive results have been obtained in recent years for the
four-nucleon low-energy interaction (see e.g.~\cite{Fons99} and
references therein), as well as for pion absorption on three-nucleon
systems~\cite{Avish83,Ueda89,Cant98}. With respect to other techniques 
we would like to refer to recent work in \cite{Carb99} and \cite{Viv01}. 

The paper is organized as follows. In the next section
we outline briefly the formal aspects concerning the application 
of the four-body formalism to the $\eta$-$3N$ system within the quasiparticle 
approach. Besides the basic equations we introduce here the two- and 
three-body ingredients of the model. Our results for $\eta$-$^3$H elastic 
scattering are presented in Sect.~\ref{sect3}, and conclusions are drawn in
the last section. Details are given in two appendices. 

\section{Application of the four-body formalism to the $\eta$-$3N$ system}
\label{sect2}

The separable method is well known to allow one 
to reduce the $n$-body problem to an in general simpler $(n-1)$-body case,
where one of the constituents appears as a quasiparticle, i.e., a two-body
bound, virtual or resonance state. In particular, the three-body equations
for the (3$\to$3) transition amplitudes are exactly reduced to effective
two-particle equations of Lippmann-Schwinger type. Their kernels contain
the off-energy-shell (2$\to$2) amplitudes for all two-body subsystems.
In an analogous manner, the four-body scattering kernels can be expressed
in terms of subamplitudes stemming from the decomposition of the four-body
system into the partitions (1+3) and (2+2). Therefore,
approximating these subamplitudes once more by a separable ansatz we can
again reduce the four-body problem to an effective quasiparticle
two-body one. The formal details of this two-step reduction scheme may
be found in Refs.~\cite{GS67,Fons86,Narod81}. Here we restrict ourselves 
to a brief description of the resulting quasi-two-body equations as
applied to the $\eta$-$3N$ system. 

\subsection{The four-body $\eta$-$3N$ equations}
\label{sect2a}
Since our formalism does not include
Coulomb forces, we will consider for definiteness 
the $\eta$-$^3$H interaction throughout this paper.
All appropriate states are assumed to be properly
antisymmetrized with respect to the nucleons. The corresponding
antisymmetrization procedure is outlined in Appendix A.
Then we are led to the following three
channels, corresponding to three possible two-quasiparticle partitions of the
$\eta$-$3N$ system 
\begin{equation}\label{eq145}
(1)\!:\ \eta+(3N)\,, \quad   (2)\!:\ N+(\eta NN)\,, \quad  (3)\!:\
(N\eta)+(NN)\,.
\end{equation}
We only need the amplitudes connecting the initial asymptotic 
state, consisting of the
$3N$ bound state ($^3$H) and a free $\eta$-meson, with all three channels
listed in (\ref{eq145}). They obey a set of three coupled integral equations
(see Appendix A), whose structure is represented by the following matrix
equation
\begin{equation}\label{eq150}
\left( \begin{array}{c} X_1 \\ X_2 \\ X_3
\end{array} \right)= \left( \begin{array}{c} 0\\ Z_{21} \\ Z_{31}
\end{array} \right)+
\left( \begin{array}{ccc}
0      & Z_{12} & Z_{13} \\
Z_{21} & Z_{22} & Z_{23} \\
Z_{31} & Z_{32} & 0      \\
\end{array}\right)
\left( \begin{array}{cccc}
 \Theta_1 & &  \\
& \Theta_2  &  \\
& & \Theta_3   \end{array}\right)
\left( \begin{array}{c} X_1 \\ X_2 \\ X_3 \end{array} \right)\,.
\end{equation}
Here the index $\alpha$=1,\,2,\,3 stands for the channel $(\alpha)$ from
(\ref{eq145}). The amplitude $X_\alpha$ describes the transition
$(1)\to(\alpha)$. The effective potentials\footnote{Following the work
\protect{\cite{GS67}} we explore the formal 
analogy with the Lippmann-Schwinger equation and use for the driving
terms $Z_{\alpha\beta}$ the suggestive term ``potential''.} 
$Z_{\alpha\beta}$ are expressed in terms 
of the form factors, generated by the separable representation of the
subamplitudes appearing in the channels (\ref{eq145}).

\begin{figure}
\centerline{\psfig{figure=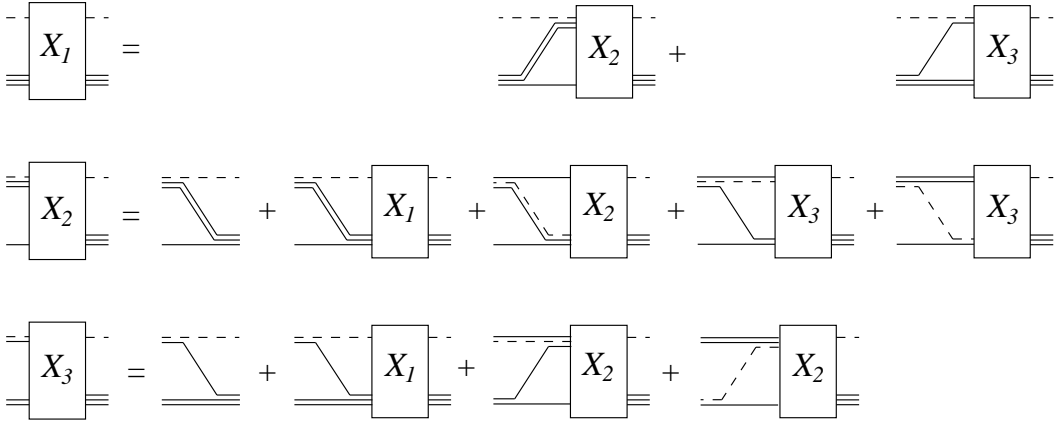,width=14cm,angle=0}}
\vspace{.5cm} \caption{ Diagrammatic representation of the
equations (\protect{\ref{eq155}}) for the transition amplitudes
$X_\alpha$ of $\eta$-$3N$ scattering.
} \label{fig1}
\end{figure}

Because here we consider
only low energy scattering, we take into account only the dominant
$s$-wave part of the interaction in the two-body subsystems and thus only
$s$-waves in the three- and four-particle states. Therefore, the matrix
elements are diagonal with respect to the total spin $S$. For the elastic
$\eta$-$^3$H scattering we need to consider only those states
where the spins and isospins of all nucleons are coupled to $S=T=1/2$.
In all expressions to follow we drop the index $L=0$.  The explicit
analytical form of the potentials $Z_{\alpha\beta}$, taking into account the
spin-isospin degrees of freedom, is given in the Appendix B. In a more
detailed notation the system (\ref{eq150}) reads
\begin{eqnarray}\label{eq155}
X_{\alpha;nn'}^{(ss')}(p,p',E)&=&
(1-\delta_{\alpha 1})Z_{\alpha1;nn'}^{(ss')}(p,p',E)\nonumber\\
&&+\sum_{\beta=1}^3\Big(1-\delta_{\alpha\beta}(1-\delta_{\alpha 2})\Big)
\sum\limits_{n''s''}
\int\limits_0^{\infty}
Z_{\alpha\beta;nn''}^{(ss'')}(p,p'',E)\Theta_{\beta;n''}({\cal E}_\beta)
  X_{\beta;n''n'}^{(s''s')}(p'',p',E)\,\frac{p''^2\,dp}{2\pi^2}\,.
\end{eqnarray}
The index $s=(0,1)$ in the above equations
corresponds to the total spin of the given $NN$ pair. Clearly,
due to the pseudoscalar-isoscalar nature of the $\eta$-meson, the value
of $s$ fixes uniquely the spin structure of the overall 4-body state with
total spin $S$=1/2. In view of the limitation of the two-body interaction
to the dominant $s$-wave part, the isospin $t$ of a $NN$ pair is fixed by
its spin $s$ through the condition $s+t=1$.
The subenergies ${\cal E}_\beta$ in (\ref{eq155}) are defined as
\begin{equation}\label{eq160}
{\cal E}_\beta=E-\frac{p''^2}{2M_\beta}\,, \ (\beta=1,2,3)\,,
\end{equation}
with reduced masses
\begin{equation}\label{eq165}
M_1=\frac{3M_Nm_\eta}{3M_N+m_\eta}, \
M_2=\frac{M_N(2M_N+m_\eta)}{3M_N+m_\eta}, \
M_3=\frac{2M_N(M_N+m_\eta)}{3M_N+m_\eta}.
\end{equation}
A graphical representation of the system (\ref{eq155}) is shown in
Figs.~\ref{fig1} and~\ref{fig2}. 

The structure of (\ref{eq150}) allows one to eliminate the channel (1)
yielding an equivalent set of only two coupled equations for the amplitudes
$X_2$ and $X_3$. In detail, one has for $\alpha\in\{2,\,3\}$
\begin{eqnarray}\label{eq175}
X_{\alpha;nn'}^{(ss')}(p,p',E)&=&
Z_{\alpha 1;nn'}^{(ss')}(p,p',E)
+ \sum_{\beta=2,3}\sum\limits_{n''s''}\int\limits_0^{\infty}\widetilde
Z_{\alpha\beta;nn''}^{(ss'')}(p,p'',E)\Theta_{\beta;n''}({\cal E}_\beta)
  X_{\beta;n''n'}^{(s''s')}(p'',p',E)\,\frac{p''^2\,dp}{2\pi^2}\,,
\end{eqnarray}
where the new effective potentials are given by
($\alpha,\,\beta\in\{2,\,3\}$)
\begin{eqnarray}\label{eq180}
\widetilde{Z}_{\alpha\beta;nn'}^{(ss')}(p,p',E)&=&
\Big(1-\delta_{\alpha\beta}(1-\delta_{\alpha 2})\Big)
Z_{\alpha\beta;nn'}^{(ss')}(p,p',E) \nonumber \\
&&+ \sum\limits_{n''s''}\int
\limits_0^{\infty}
Z_{\alpha 1;nn''}^{(ss'')}(p,p'',E)\Theta_{1;n''}({\cal E}_1)
Z_{1\beta;n''n'}^{(s''s')}(p'',p',E)\frac{p''^2dp''}{2\pi^2}\,.
\end{eqnarray}
After solving the system (\ref{eq175}), the amplitude $X_1$ is obtained from
\begin{eqnarray}\label{eq185}
X_{1;nn'}^{(ss')}(p,p',E)&=&
 \sum_{\beta=2,3}\sum\limits_{n''s''}\int\limits_0^{\infty}
Z_{1\beta;nn''}^{(ss'')}(p,p'',E)\Theta_{\beta;n''}({\cal E}_\beta)
  X_{\beta;n''n'}^{(s''s')}(p'',p',E)\,\frac{p''^2dp}{2\pi^2}\,.
\end{eqnarray}
The set of equations (\ref{eq175}) is more suitable for the numerical
solution than (\ref{eq155}), since in the former case the integration over
the triton pole in the propagator $\Theta_1$ may be carried out
independently from the procedure of solving the system (\ref{eq175})
itself. Thus the kernels in (\ref{eq175}) are smooth
functions of the integration variable, and the equations may be solved by
direct matrix inversion. We recall that below the triton break-up threshold
we are dealing with only nonsingular potentials $Z_{\alpha\beta}$.


\subsection{The subamplitudes}
\label{sect2b}
The key ingredient of the quasiparticle method \cite
{GS67,Narod81}, leading to the equations of the type (\ref{eq155}), is
the separable representation of the  
off-shell scattering amplitudes for the two- and
three-body subsystems, appearing in the (2+2)- and (1+3)-partitions.
In our case, the two types of two-body subsystems involved are
$NN$ and $\eta N$, denoted in the
following by ``$d\,$'' and ``$N^*$'', respectively.
For the corresponding scattering matrices we adopt the simplest rank-one 
separable form.
In detail, we use for the $NN$ interaction
\begin{equation}\label{eq10}
t_{d}^{(s)}(q,q',z)=g_d^{(s)}(q)\tau_d^{(s)}(z)g_d^{(s)}(q')\,,
\end{equation}
with
\begin{equation}\label{eq15}
\tau_d^{(s)}(z)=-\frac{1}{2M_N}\,\Big[
1-\frac{1}{4\pi^2}\int\limits_0^\infty
\frac{[g^{(s)}_d(q)]^2}{zM_N-q^2}\,q^2dq\Big]^{-1}\,.
\end{equation}
Here the upper index $(s)$ stands for the singlet $(s=0)$ and triplet
$(s=1)$ $NN$ states. For the form factors $g_d^{(s)}(q)$ we use the simplest
parameterization
\begin{equation}\label{eq20}
g_d^{(s)}(q)=g_d^{(s)}\frac{\beta^2_s}{\beta^2_s+q^2}\,.
\end{equation}
Analogously, we choose in the $\eta N$ channel
\begin{equation}\label{eq25}
t_{N^*}(q,q',z)=g_{N^*}^{\eta}(q)\tau_{N^*}(z)g_{N^*}^{\eta}(q')\,,
\end{equation}
with
\begin{equation}\label{eq30}
\tau_{N^*}(z)=-\Big[z+M_N+m_\eta-M_0-\Sigma_\eta(z)-\Sigma_\pi(z)\Big]^{-1}\,,
\end{equation}
where
\begin{equation}\label{eq35}
\Sigma_\rho (z)=\int\limits_0^{\infty}\frac{[g_{N^*}^\rho (q)]^2}
{z+m_\eta-m_\rho -\frac{q^2}{2\mu_{\rho N}}+i\epsilon}\,\frac{q^2dq}{2\pi^2}\,,
\end{equation}
and
\begin{equation}\label{eq40}
g_{N^*}^\rho (q)=g_{N^*}^\rho \frac{\beta^2_\rho }{\beta^2_\rho +q^2}\,,\ \
\mu_{\rho N}=\frac{M_Nm_\rho }{M_N+m_\rho }\,,\
\quad \rho \in\{\pi,\eta\}\,.\nonumber
\end{equation}
The parameters appearing in (\ref{eq20}), (\ref{eq30}) and (\ref{eq40}) are
summarized in Table~\ref{tab1}. Those for the $NN$ interaction were
taken from the low-energy $NN$ fit presented in \cite{Yam54}, and the $\eta
N$-parameters are chosen as to reproduce the scattering length $a_{\eta
N}=(0.75+i\,0.27)$~fm which agrees with the most recent
results~\cite{Batin95,Wyce97}. However, our parameterization gives a
different value for the effective range parameter
$r_0=(1.95+i\,0.07)$~fm compared to $r_0=(1.5+i\,0.24)$~fm of~\cite{Wyce97}.
Unfortunately, we are not able to fit both values $a_{\eta N}$ and
$r_0$ simultaneously, which is of course the price one has to pay
for using the simplest separable ansatz (\ref{eq10}). When comparing
our predictions with those of~\cite{Wyce95,Belya95} we use also other sets of
$\eta N$-parameters chosen in such a way that they lead in each case
to the corresponding value of $a_{\eta N}$ used there
(see Table~\ref{tab3}).  

Turning now to the general scheme, we have to introduce also the
separable representation for the three-body subamplitudes,
which will then serve as a necessary input for the four-body
calculation. For this purpose we apply the Hilbert-Schmidt expansion. The
main formal aspects of the procedure can be found e.g.\ in
\cite{Wein64,Sitenko}. 

The three-nucleon $s$-wave doublet amplitudes $U_{1;ss'}$, appearing in
the channel 
(1) obey the equation (see e.g.~\cite{Love64})
\begin{equation}\label{eq45}
U_{1;ss'}(q,q',{\cal E})=V_{1;ss'}(q,q',{\cal E}) + \!\sum\limits_{s''=0,1}
\int\limits_0^{\infty}V_{1;ss''}(q,q'',{\cal E})
\,\tau_d^{(s'')}\!\Big({\cal
E}-\frac{3q''^2}{4M_N}\Big)U_{1;s''s'}(q'',q',{\cal E})
\frac{q''^2dq''}{2\pi^2}\,,\quad s,s'=0,1\,.
\end{equation}
The effective potentials are defined in terms of the form factors
$g_d(k)$ of the two-nucleon amplitude (\ref{eq10}) as
\begin{equation}\label{eq50}
V_{1;ss'}(q,q',{\cal
E})=\frac{\Lambda_{ss'}}{2}\int\limits_{-1}^{+1}
\frac{g_d^{(s)}(|\vec{q}\,'+\frac{1}{2}\vec{q}\,|)\ g_d^{(s')}
(|\vec{q}+\frac{1}{2}\vec{q}\,'|)}{{\cal E}-\frac{q^2}{M_N}-\frac{q'^2}
{M_N}-\frac{\vec{q}\cdot\vec{q}\,'}{M_N}}
d(\hat{q}\cdot\hat{q}')\,, \ \quad s,s'=0,1\,,
\end{equation}
where ${\cal E}$ denotes the total c.m.\ kinetic energy in the three
nucleon system, and the matrix of the spin-isospin coefficients is given by
\begin{equation}\label{eq55}
\Lambda=\left(
\begin{array}{rr}
\frac{1}{2} & -\frac{3}{2} \\
 & \\
-\frac{3}{2} & \frac{1}{2}  \end{array} \right)\,.
\end{equation}
The Hilbert-Schmidt expansion for the driving term $V_{1;ss'}$ reads
\begin{equation}\label{eq80}
V_{1;ss'}(q,q',{\cal E})=-\sum\limits_{n} \lambda_{n}({\cal E})
u^{(s)}_{n}(q,{\cal E})u^{(s')}_{n}(q',{\cal E})\,,
\end{equation}
where the functions $u_{n}^{(s)}(q,{\cal E})$ are taken as the
eigenfunctions of the kernel of equation (\ref{eq45}) with the eigenvalues
$\lambda_{n}$, i.e.
\begin{equation}\label{eq85}
u^{(s)}_{n}(q,{\cal E})=\frac{1}{\lambda_{n}({\cal E})}\sum\limits_{s'=0,1}
\int\limits_0^{\infty} V_{1;ss'}(q,q',{\cal E})
\,\tau^{(s')}_d\!\Big({\cal E}-\frac{3q'^2}{4M_N}\Big)
u^{(s')}_{n}(q',{\cal E})
\frac{q'^2dq'}{2\pi^2}\,,
\ \quad s=0,1\,.
\end{equation}
They are normalized according to
\begin{equation}\label{eq90}
\sum\limits_{s=0,1}\int\limits_0^{\infty}u^{(s)}_{n}(q,{\cal E})
\,\tau^{(s)}_d\!\Big({\cal E}-\frac{3q^2}{4M_N}\Big)
u^{(s)}_{n'}(q,{\cal E})
\frac{q^2dq}{2\pi^2}=-\delta_{nn'}\,.
\end{equation}
The separable form of the amplitude can easily be found
\begin{equation}\label{eq95}
U_{1;ss'}(q,q',{\cal E})=\sum\limits_{n}
u^{(s)}_{n}(q,{\cal E})\Theta_{1;n}({\cal E})u^{(s')}_{n}(q',{\cal E})\,, \
\quad \Theta_{1;n}({\cal E})=
\frac{\lambda_{n}({\cal E})}{\lambda_{n}({\cal E})-1}\,.
\end{equation}

The amplitudes for the $\eta NN$ scattering, related to the channel
(2) and denoted in the following by $\overline{U}_{2;ij}$ with
$i,j\in\{d,N^*\}$, are coupled into two independent sets corresponding
to two possible 
spin-isospin $\eta NN$ $s$-wave states $(S;T)=(0;1)$ and $(1;0)$, denoted by 
an upper index $(s)$ with $s\in\{0,1\}$, each of the form
\begin{eqnarray}\label{eq60}
\overline{U}^{(s)}_{2;dd}(q,q',{\cal E})&=&\int\limits_0^{\infty}
V^{(s)}_{2;dN^*}(q,q'',{\cal E}) \,\tau_{N^*}\!\Big({\cal
E}-\frac{q''^2}{2\widetilde{\mu}_{N^*}}\Big)
\overline{U}^{(s)}_{2;N^*d}(q'',q',{\cal
E})\frac{q''^2dq''}{2\pi^2}\,,\\
\overline{U}^{(s)}_{2;N^*d}(q,q',{\cal E})&=&
V^{(s)}_{2;N^*d}(q,q',{\cal E})+ \int\limits_0^{\infty} \Big[
V^{(s)}_{2;N^*d}(q,q'',{\cal E}) \,\tau^{(s)}_d\!\Big({\cal
E}-\frac{q''^2}{2\widetilde{\mu}_d}\Big)
\overline{U}^{(s)}_{2;dd}(q'',q',{\cal E})
\nonumber\\
&&\hspace*{2cm}+V^{(s)}_{2;N^*N^*}(q,q'',{\cal E})
\,\tau_{N^*}\!\Big({\cal E}-\frac{q''^2}{2\widetilde{\mu}_{N^*}}\Big)
\overline{U}^{(s)}_{2;N^*d}(q'',q',{\cal E})\Big]
\frac{q''^2dq''}{2\pi^2}\,.
\nonumber
\end{eqnarray}
The reduced masses appearing in (\ref{eq60}) are defined by
\begin{equation}\label{eq65}
\widetilde{\mu}_{N^*}=\frac{M_N(M_N+m_\eta)}{2M_N+m_\eta},\ \
\widetilde{\mu}_d=\frac{2M_Nm_\eta}{2M_N+m_\eta}\,.
\end{equation}
The corresponding effective potentials are
\begin{eqnarray}\label{eq70}
&&V^{(s)}_{2;dd}(q,q',{\cal E})=0\,, \nonumber \\
&&V^{(s)}_{2;dN^*}(q,q',{\cal E})=\frac{1}{\sqrt{2}}\int\limits_{-1}^{+1}
\frac{g^{(s)}_d(|\vec{q}\,'+\frac{1}{2}\vec{q}\,|)
\ g_{N^*}^{\eta}(|\vec{q}+\frac{\mu_{\eta N}}{M_N}\vec{q}\,'|)}
{{\cal E}-\frac{q^2}{M_N}-\frac{q'^2}{2\mu_{\eta
N}}-\frac{\vec{q}\cdot\vec{q}\,'} {M_N}}d(\hat{q}\cdot\hat{q}')
\,,\nonumber\\
&&V^{(s)}_{2;N^*d}(q,q',{\cal E})= V^{(s)}_{2;dN^*}(q',q,{\cal
E})\,, \\
&&V^{(s)}_{2;N^*N^*}(q,q',{\cal E})=\frac{1}{2}\int\limits_{-1}^{+1}
\frac{g^{\eta}_{N^*}(|\vec{q}\,'+\frac{\mu_{\eta N}}{M_N}\vec{q}\,|)
\ g_{N^*}^{\eta}(|\vec{q}+\frac{\mu_{\eta N}}{M_N}\vec{q}\,'|)}
{{\cal E}-\frac{q^2}{M_N}-\frac{q'^2}{2\mu_{\eta N}}
-\frac{\vec{q}\cdot\vec{q}\,'}{m_\eta}} d(\hat{q}\cdot\hat{q}')\,,
\nonumber
\end{eqnarray}
where ${\cal E}$ is the total c.m.\ kinetic energy of the $\eta NN$ system.
The properly antisymmetrized amplitudes for
$\eta NN$ scattering are expressed in terms of $\overline{U}^{(s)}_{2;ij}$ as
(see \cite{Shev98,FiAr01})
\begin{equation}\label{eq75}
U^{(s)}_{2;dd}=\overline{U}^{(s)}_{2;dd}\,, \qquad
U^{(s)}_{2;N^*d}=\frac{1}{\sqrt{2}}\overline{U}^{(s)}_{2;N^*d}\,,
\qquad s=0,1\,.
\end{equation}
Analogously to (\ref{eq95}) we have $(i,j\in\{d,N^*\})$
\begin{equation}\label{eq100}
U^{(s)}_{2;ij}(q,q',{\cal E})=\sum\limits_{n}
v^{(s)}_{i;n}(q,{\cal E})\Theta^{(s)}_{2;n}({\cal E})
v^{(s)}_{j;n}(q',{\cal E})\,, \ \quad
\Theta^{(s)}_{2;n}({\cal E})
=\frac{\eta^{(s)}_{n}({\cal E})}{\eta^{(s)}_{n}({\cal E})-1}\,,
\end{equation}
where the form factors $v^{(s)}_{i;n}$ obey the homogeneous equation
\begin{equation}\label{eq105}
v^{(s)}_{i;n}(q,{\cal E})
=\frac{1}{\eta^{(s)}_{n}({\cal E})} \sum\limits_{j=d,N^*} \int
\limits_0^{\infty}
V^{(s)}_{2;ij}(q,q',{\cal E}) \,\tau^{(s)}_j\!\Big({\cal E}
-\frac{q^{'2}}{2\widetilde{\mu}_j}\Big)
v^{(s)}_{j;n}(q',{\cal E})
\frac{q'^2dq'}{2\pi^2}\,,
\ \quad s=0,1\,.
\end{equation}
Here, of course, one has $\tau^{(s)}_{N^*}=\tau_{N^*}$ $(s=0,1)$,
and the eigenfunctions are normalized as
\begin{equation}\label{eq110}
\sum\limits_{i=d,N^*}\int\limits_0^{\infty}
v_{i;n}^{(s)}(q,{\cal E})
\,\tau_{i}^{(s)}\!\Big({\cal
E}-\frac{q^2}{2\widetilde{\mu}_i}\Big)
v_{i;n'}^{(s)}(q,{\cal E})
\frac{q^2dq}{2\pi^2}=-\delta_{nn'}\,.
\end{equation}
In the actual calculation we have neglected any $\pi NN$
states. Their inclusion would imply an increase in the number
of channels in the final equations as well as adoption of relativistic
kinematics which would lead to much more complex formalism.
On the other hand due to its small mass, the pion is
expected to give only minor corrections to low-energy
$\eta$-nucleus scattering \cite{FiAr01,Wyce01}.

Apart from the genuine three particle scattering amplitudes,
we also need as input the effective amplitudes, denoted here by
$U^{(s)}_{3;ij}$, which describe two independent pairs of interacting
particles in the channel (3). The corresponding equations read in our case
\begin{eqnarray} \label{eq125}
U^{(s)}_{3;dd}(q,q',{\cal
E})&=&\int\limits_0^{\infty} V^{(s)}_{3;dN^*}(q,q'',{\cal E})
\,\tau_{N^*}\!\Big({\cal E}-\frac{q''^2}{2\nu_{N^*}}\Big)
V^{(s)}_{3;N^*d}(q'',q',{\cal E})\frac{q''^2dq''}{2\pi^2}\,,\\
U^{(s)}_{3;N^*d}(q,q',{\cal E})&=&V^{(s)}_{3;N^*d}(q,q',{\cal E})+
\int\limits_0^{\infty}
V^{(s)}_{3;N^*d}(q,q'',{\cal E})
\,\tau^{(s)}_d\!\Big({\cal E}-\frac{q''^2}{2\nu_d}\Big)
U^{(s)}_{3;dd}(q'',q',{\cal E})\frac{q''^2dq''}{2\pi^2}\,,\nonumber
\end{eqnarray}
where ${\cal E}$ is the sum of the internal energies in the
$NN$  and $\eta N$ subsystems.
In the expressions (\ref{eq125}) the notations
$\nu_d=\mu_{\eta N}$ and $\nu_{N^*}=M_N/2$ are used.
The effective potentials are
\begin{eqnarray}\label{eq130}
&&V^{(s)}_{3;dd}(q,q',{\cal E})=0\,,\nonumber \\
&&V^{(s)}_{3;dN^*}(q,q',{\cal E})=
\frac{g_d^{(s)}(q')\ g_{N^*}^{\eta}(q)}
{{\cal E}-\frac{q^2}{2\mu_{\eta N}}-\frac{q'^2}{M_N}}\,,\\
&&V^{(s)}_{3;N^*d}(q,q',{\cal E})=
V^{(s)}_{3;dN^*}(q',q,{\cal E})\,,\nonumber \\
&&V^{(s)}_{3;N^*N^*}(q,q',{\cal E})=0\,.\nonumber
\end{eqnarray}
Analogously to the treatment above, we introduce the form factors
$w_{i;n}^{(s)}$ ($i\in\{d,N^*\}$) as the eigenfunctions of the
Lippmann-Schwinger kernel
\begin{equation}\label{eq135}
w^{(s)}_{i;n}(q,{\cal E})=\frac{1}{\xi^{(s)}_{n}({\cal E})}
\sum\limits_{j=d,N^*}\int\limits_0^{\infty}
V^{(s)}_{3;ij}(q,q',{\cal E}) \,\tau^{(s)}_j\!\Big({\cal
E}-\frac{q'^2}{2\nu_j}\Big) w^{(s)}_{j;n}(q',{\cal
E})\frac{q'^2dq'}{2\pi^2}\,, \quad i,j\in\{d,N^*\}\,,
\end{equation}
with an orthogonality condition analogous to (\ref{eq110}).
Then the separable form of the transition matrices is
generated by the Hilbert-Schmidt expansion
\begin{equation}\label{eq140}
U^{(s)}_{3;ij}(q,q',{\cal E})=\sum\limits_{n}
w_{i;n}^{(s)}(q,{\cal E})\Theta_{3;n}^{(s)}({\cal E})w_{j;n}^{(s)}
(q',{\cal E})\,, \ \quad
\Theta^{(s)}_{3;n}({\cal E})=\frac{\xi^{(s)}_{n}({\cal
E})}{\xi^{(s)}_{n}({\cal E})-1}\,.
\end{equation}
In the spirit of the terminology, adopted for the separable approach, we may
interpret the function $w_{i;n}$ as the form factor of the
two-particle ``bound state'' $j$ when the other two particles are
in the ``bound state'' $i$.

In Figs.\,\ref{fig3} through \ref{fig5}
we present the leading eigenvalues $\lambda_{n}({\cal E})$ and the
real parts of $\eta_{n}^{(s)}({\cal E})$ and
$\xi_{n}^{(s)}({\cal E})$. Several comments are in order:

1) The validity of separable expansion above is
strongly limited to the energy region below the triton break-up
threshold $^3$H~$\to n+d$,
i.e., to energies ${\cal E}<\varepsilon_{d}$,
where $\varepsilon_{d}\approx -2.22$~MeV denotes the deuteron binding energy.
Above this region (${\cal E}\ge\varepsilon_{d}$),
due to the singularities appearing in the propagators as well as in the
potentials, the kernels of the equations become noncompact,
and the Hilbert-Schmidt expansion loses its meaning.
Therefore we shall restrict our consideration to
the region below the triton break-up threshold.

2) The eigenvalue $\lambda_{1}^{(1)}$ goes through unity at the energy
${\cal E}={\cal E}_{^3\mathrm{H}}\approx -13$ MeV, which is essentially lower
than the experimental value of the triton binding energy
${\cal E}_{^3\mathrm{H}}^{\mathrm{exp}}\approx-8.5$~MeV.
This disagreement is a consequence of using the extremely simple
Yamaguchi $NN$ scattering matrix (\ref{eq10}). Although this ansatz fits
low-energy $NN$ scattering, it has too long a tail into the high-momentum
region and, therefore, predicts too much binding for the three-nucleon
system.

3) The modulus of the eigenvalues $\lambda_{i}$ and $\lambda_{i+1}$
are close to each other. Therefore, one can expect, that the
leading terms in the expansion (\ref{eq95}) will cancel each other to a
large extent, which may lead to a relatively slow
convergence rate of the sum.  In order to demonstrate how well a
finite sum (\ref{eq95}) may represent the exact amplitude, we show in
Fig.~\ref{fig3} the ratio of the Schmidt norms
\begin{equation}\label{eq115}
R_n({\cal E})=\frac{\|{\cal O}_n\|}{\|{\cal O}\|}\,,
\end{equation}
of the operators
\begin{equation}\label{eq120}
{\cal O}=\sqrt{\tau_d^{(1)}}V^{N}_{11}\sqrt{\tau_d^{(1)}}\,,\ \
{\cal O}_n=\sqrt{\tau_d^{(1)}}(V^{N}_{11}-V^{N}_{11,n})
\sqrt{\tau_d^{(1)}}\,,
\end{equation}
where $V^{N}_{11,n}$ is given
by the sum (\ref{eq80}) containing only the first $n$ terms.
We see that the rate of convergence is not very impressive but appears
to be sufficient for
the practical calculation.  Already in the region close to the two-body
threshold (${\cal E}=\varepsilon_{n,d}$) the six leading terms in (\ref{eq80})
accumulate more than 95~\% of the Schmidt norm of ${\cal O}$.
The same is valid for the $\eta NN$ case and especially for the $(\eta
N)+(NN)$ channel as may be seen from Figs.~\ref{fig4} and \ref{fig5}.

The triton wave function was extracted from the pole of the $3N$
scattering amplitude. In the present calculation we have
restricted ourselves to the principal, spatially completely
symmetric $s$-wave part, which in the usual notation (see
e.g.~\cite{Schiff64}) reads
\begin{equation}\label{eq190}
\Psi_{^3\mathrm{H}}=-\xi^a\Psi^s(\vec{p}_1,\vec{p}_2,\vec{p}_3)\,,
\end{equation}
where $\xi^a$ denotes the completely antisymmetric spin-isospin part,
given by
\begin{equation}\label{eq195}
\xi^a=\frac{1}{\sqrt{2}}(\chi^{(0)}\zeta^{(1)}-\chi^{(1)}\zeta^{(0)})\,.
\end{equation}
The spin function $\chi^{(s)}$ ($s=0,1$) describes a three-nucleon spin=1/2
state which is obtained by coupling first the spins of two nucleons to a
spin $s$ and then by coupling $s$ with the spin of the third nucleon to a
total spin 1/2. The isospin function $\zeta^{(t)}$ ($t=0,1$) is constructed
in the same manner.

The spatial part $\Psi^s$, completely
symmetrical with respect to the nucleon momenta, is given by
\begin{equation}\label{eq200}
\Psi^s(\vec{p}_1,\vec{p}_2,\vec{p}_3)=\frac{1}{\sqrt{3}}
(1+P_{13}+P_{23})\psi(k_{12},q_3)\,.
\end{equation}
Here, $k_{12}$ and $q_3$ are the usual Jacobi momenta for the tree (1+2)+3.
The function $\psi(k,q)$ can be expressed within our formalism in
terms of the eigenfunctions (\ref{eq85}) as follows
\begin{equation}\label{eq205}
\psi(k,q)=\frac{1}{\sqrt{2}}\Big(u^{(0)}(k,q)-u^{(1)}(k,q)\Big)\,,
\end{equation}
with
\begin{equation}\label{eq210}
u^{(s)}(k,q)=N\,g_d^{(s)}(k)\,\tau_d^{(s)}\!
\Big({\cal E}_{^3\mathrm{H}}-3q^2/4M_N\Big)
\frac{u_1^{(s)}(q,{\cal E}_{^3\mathrm{H}})}
{{\cal E}_{^3\mathrm{H}}-3q^2/4M_N-k^2/M_N}\,.
\end{equation}
The normalization factor $N$ is taken from the residue of the $3N$ scattering
matrix (7) at ${\cal E}={\cal E}_{^3\mathrm{H}}$, and one finds
\begin{equation}\label{eq215}
N^{-2}=\Big[\frac{d\lambda_1}{d\cal{E}}\Big]_{{\cal E}_{^3\mathrm{H}}}\,.
\end{equation}

The elastic $\eta$-$^3$H scattering amplitude $F_{\eta\, ^3\mathrm{H}}(p)$
is then expressed in terms of the amplitudes
$X_{1;11}^{(ss')}$, taken on the energy shell,
\begin{equation}\label{eq220}
F_{\eta\,^3\mathrm{H}}(p)=-\frac{\mu_{\eta\,^3\mathrm{H}}}{2\pi}
\Big(X_{1;11}^{(00)}(p,p,E)+
X_{1;11}^{(11)}(p,p,E)-2X_{1;11}^{(10)}(p,p,E) \Big)\,
\end{equation}
with $p=\sqrt{2\mu_{\eta\,^3\mathrm{H}}E}$ and $\mu_{\eta\,^3\mathrm{H}}$
being
the $\eta$-$^3$H reduced mass. One should note that in the calculation of
the amplitudes $X_1$ entering (\ref{eq220}), the driving
terms $Z_{1i}$ and $Z_{i1}$ ($i=2,3$) in (\ref{eq175}) and (\ref{eq185})
were redefined in view of the approximation (\ref{eq190}) (see Appendix B).


\section{Results and discussion}
\label{sect3}

In Table~\ref{tab2} we present our results for the $\eta$-$^3$H
scattering length obtained by keeping a finite number of terms in the
Hilbert-Schmidt expansion of the amplitudes (\ref{eq95}), (\ref{eq100})
and (\ref{eq140}). One can see, 
that the rate of convergence of $a_{\eta\,^3\mathrm{H}}$ is
approximately the same as that for the integral kernels discussed in
the previous sections (see Figs.~\ref{fig3}, \ref{fig4} and \ref{fig5}).
The choice $n_1=4, n_2=n_3=6$
provides rather satisfactory accuracy. Our result obtained with the
$\eta N$-input corresponding to $a_{\eta N}=(0.75+i\,0.27)$~fm
\cite{Wyce97} is 
\begin{equation}\label{eq225}
a_{\eta\,^3\mathrm{H}}=(4.2+i\,5.7)\,\mbox{fm}\,.
\end{equation}
Firstly we note that the
scattering length is rather large, which may be explained as a
consequence of the virtual state (the real part of the scattering
length is positive) generated by the strong attraction of the
$\eta$-$3N$ interaction. The large magnitude of
$a_{\eta\,^3\mathrm{H}}$ indicates that the corresponding pole of the
scattering amplitude $F_{\eta\,^3\mathrm{H}}(p)$
lies near the elastic scattering threshold. Turning to the region of positive
energies, we present in Fig.~\ref{fig6} the total cross section for
elastic $\eta$-$^3$H scattering.  Here the virtual $\eta$-$^3$H state manifests
itself in a strong enhancement of $\sigma(E)$ when $E$ approaches the
threshold value. It would appear natural that this effect is really
observed, e.g., in $\eta$-production in $pd$-collisions \cite{May96}.

As already mentioned, we compare in Table~\ref{tab3} our predictions
on the $\eta$-$^3$H scattering length with those
of~\cite{Wyce95,Belya95} using an $\eta N$-interaction which reproduces
the $a_{\eta N}$ scattering length of~\cite{Wyce95,Belya95}. One must 
note a rather 
strong difference of the results, especially for the real part of
$a_{\eta\,^3\mathrm{H}}$. Apparently, the latter must be very
sensitive to the position
of the pole in the amplitude as it lies close to zero energy. Very
rough agreement may be noticed with~\cite{Wyce95}. But there is a
tendency of our prediction to give a larger value of
$\Re e\,a_{\eta\,^3\mathrm{H}}$,
which is probably due to the difference in the $\eta N$ range parameter $r_0$
noted in the beginning of Sect.~\ref{sect2b}.

As to the question about the existence of a
quasibound state, which an $\eta$-meson can form with the lightest nuclei
\cite{Wilkin93,Wyce95,Belya95,Tryas97}, we see that within our model the
$\eta$-$3N$ interaction is not strong enough in order to bind the $\eta$-$^3$H
system. This conclusion does not support the results of the 
FRA-model~\cite{Belya95} or those of the optical model~\cite{Wilkin93,Tryas97}
where an $\eta$-$3N$ bound state appears already for rather modest values
of the $\eta N$ scattering length. 
It should also be remembered that our calculations are
based on the rank-one separable $NN$ potential which is known to
overestimate the attraction in the three-nucleon system. Therefore, we
expect that the use of more refined $NN$ models will most likely reduce the
probability for binding the $\eta$-$^3$H-system.


\section{Conclusion}
\label{concl}

In the present paper, the four-body scattering formalism has been applied to
study the $\eta$-$3N$ interaction in the energy region very close to
the $\eta$-$^3$H elastic scattering threshold.
The calculational scheme, which formally allows an exact solution, 
is based on the separable approximation of the appropriate
integral kernels. The validity of this approach is justified by the
fact that the driving $NN$ and $\eta N$ interactions are governed mainly by the
$S$-matrix poles, lying in each case near the low-energy region. Therefore, the
separable potentials, giving the correct structure of the amplitude close
to the poles, are expected to provide an adequate approximation.
In the present paper we have realized one of the possible
schemes for solving the $\eta$-$3N$ four-body problem. To test its
applicability, we have investigated the Hilbert-Schmidt procedure for
constructing a separable representation of finite rank of the four-body
kernels. The examination of
their Schmidt norms shows that the rate of convergence by increasing the rank
is not very fast, but satisfactory for practical purposes.
Keeping 6 to 8 terms in the separable expansion, we have obtained a rather good
precision. The same conclusion is valid for the calculation of the 
$\eta$-$^3$H scattering length. At the same time we would like to point out
that the present
calculation suffers from the oversimplification of the $NN$ interaction,
for which the rank-one Yamaguchi potential has been used. Of course,
this shortcoming can be cured by using a more sophisticated
separable approximation for the $NN$ scattering matrix, requiring only
more computational efforts.  Thus, our results may be considered as a
starting point for more realistic calculations.

\renewcommand{\theequation}{A\arabic{equation}}
\setcounter{equation}{0}
\section*{Appendix A}
\label{appA}

\begin{figure}
\centerline{\psfig{figure=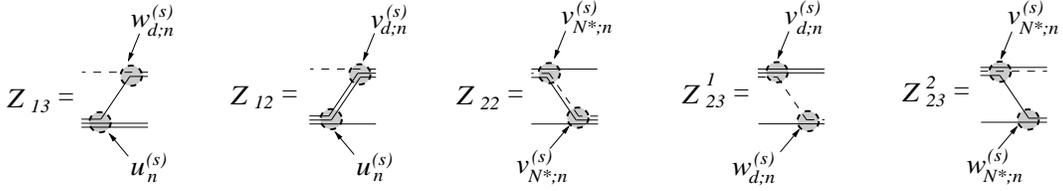,width=14cm,angle=0}}
\vspace{.5cm} \caption{ Diagrammatic representation of the potentials
$Z_{\alpha\beta}$ in the separable approximation (see Appendix B).
} \label{fig2}
\end{figure}

\begin{figure}
\centerline{\psfig{figure=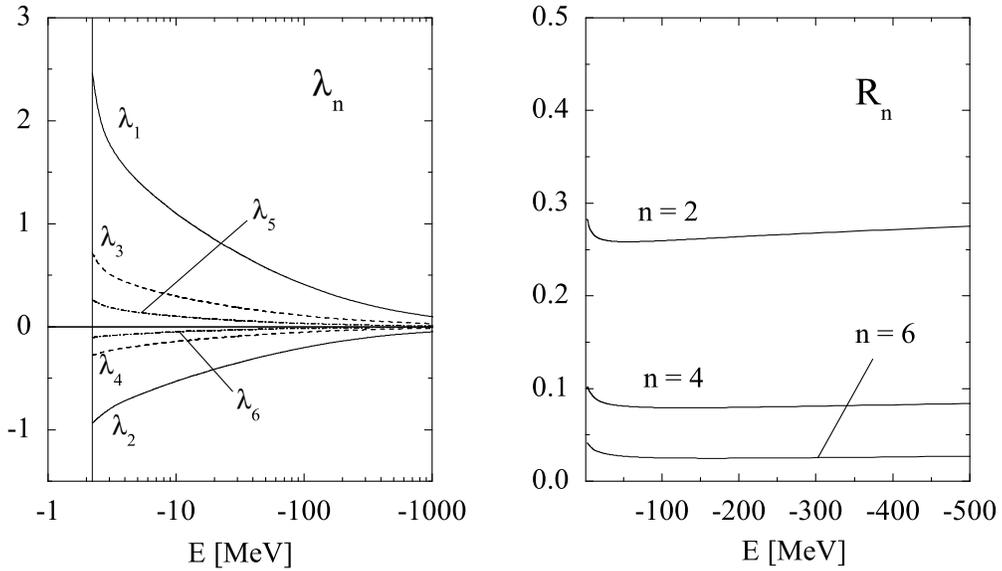,width=15cm,angle=0}}
\vspace{.5cm}
\caption{Left panel:
The leading eigenvalues of the Hilbert-Schmidt expansion for the
three-nucleon scattering amplitude in the
$S=T=1/2$ state.  Right panel: The ratio between the Schmidt norms
of the kernels ${\cal O}$ and ${\cal O}_n$ as defined by
(\protect{\ref{eq115}}) and (\protect{\ref{eq120}}).
}
\label{fig3}
\end{figure}

\begin{figure}
\centerline{\psfig{figure=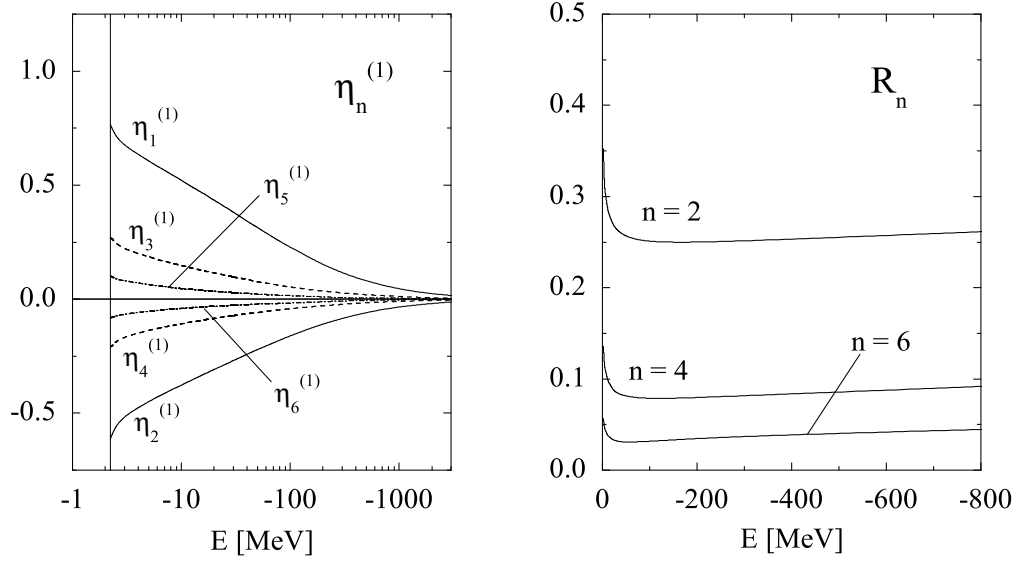,width=15cm,angle=0}}
\vspace{.5cm}
\caption{The same as in Fig.~\protect{\ref{fig3}} for
$\eta NN$-scattering in the $(S;T)=(1;0)$ state.
}
\label{fig4}
\end{figure}

\begin{figure}
\centerline{\psfig{figure=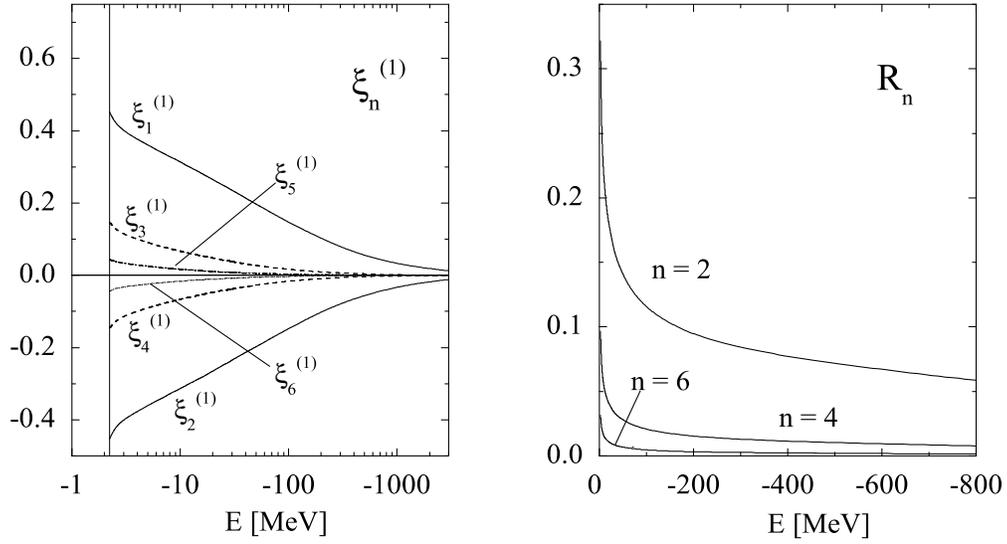,width=15cm,angle=0}}
\vspace{.5cm}
\caption{The same as in Fig.~\protect{\ref{fig3}} for the
$(\eta N)$+$(NN)$ partition.  }
\label{fig5}
\end{figure}

\begin{figure}
\centerline{\psfig{figure=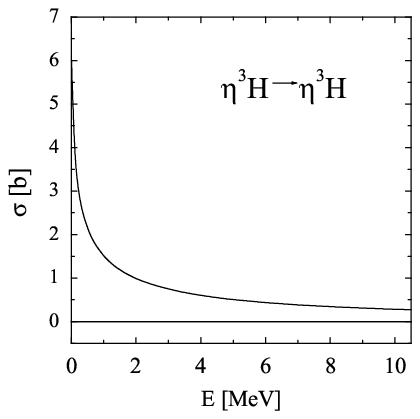,width=9cm,angle=0}}
\vspace{.5cm} \caption{ Total cross section of elastic
$\eta$-$^3$H-scattering versus the c.m.\ kinetic energy $E$. }
\label{fig6}
\end{figure}

In this appendix the transition amplitudes for $\eta$-$3N$ scattering
are defined with due consideration of the Pauli principle for
nucleons. We begin with the two-body equations resulting from
the quasiparticle approach to the four-body problem~\cite{GS67,Narod81}
\begin{equation}\label{A160}
X_{\alpha\beta; nn'}(\vec{p},\vec{p}\,'\!,E) =
Z_{\alpha\beta;nn'}(\vec{p},\vec{p}\,'\!,E)  
+ \sum_{\gamma\ne\alpha}
\sum\limits_{n''}\int
Z_{\alpha\gamma;nn''}(\vec{p},\vec{p}\,''\!,E)
\Theta_{\gamma n''}(E) X_{\gamma\beta;n''n'}(\vec{p}\,'',\vec{p}\,'\!,E)
\frac{d^3p''}{(2\pi)^3}\,.
\end{equation}
Here the effective potentials 
\begin{equation}\label{A170}
Z_{\alpha\beta; nn'}(\vec{p},\vec{p}\,'\!,E)
=\sum\limits_i
u^\alpha_{i;n}(\vec{q},{\cal E})\tau_i(z)
u^\beta_{i;n'}(\vec{q}\,'\!,{\cal E}')
\end{equation}
are represented by the particle exchange diagrams (see e.g.\
Fig.~\ref{fig2}). They include the propagator $\tau_i(z)$ of the 
quasiparticle ``$i$'' depending on the two-body subenergy
\begin{equation}\label{A174}
z=E-\frac{p^2}{2M_\alpha}-\frac{p'^2}{2M_\beta}-\frac{\vec{p}\cdot\vec{p}\,'}
{m_k}\,,
\end{equation}
where $M_{\alpha}$ denotes the reduced mass in the two-particle 
channel $\alpha$.
The form factors $u^\alpha_{i;n}$ are generated by the separable expansion 
of the subamplitudes in the (1+3)- and (2+2)-partitions 
\begin{equation}\label{A150}
U^\alpha_{ij}(\vec{q},\vec{q}\,'\!,{\cal E})
=\sum\limits_n u^\alpha_{i;n}(\vec{q},{\cal E})
\Theta_{\alpha n}({\cal E}) u^\alpha_{j;n}(\vec{q}\,'\!,{\cal E})\,.
\end{equation}
In the case $\alpha$ = (1+3) the functions
$U^\alpha_{ij}(\vec{q},\vec{q}\,'\!,{\cal E})$  are the familiar
Lovelace amplitudes appearing within the pure separable approach to
the three-body scattering~\cite{Love64}. For $\alpha$ = (2+2) these functions
describe two independently propagating pairs of particles, each
treated as a quasiparticle. 
In (\ref{A170}) the energies ${\cal E}$ and ${\cal E}'$ in the subsystems 
$\alpha$ and $\beta$, respectively, are defined by 
\begin{equation}
{\cal E}=E-\frac{p^2}{2M_{\alpha}}
\quad \mbox{and}\quad
{\cal E}'=E-\frac{p'^2}{2M_{\beta}}\,.
\end{equation}
The momenta $\vec{q}$ and $\vec{q}\,'$ in (\ref{A170}) are of course
functions of $\vec{p}$ and $\vec{p}\,'$
\begin{equation}\label{A172}
\vec{q}=\vec{p}\,'+\frac{M_\alpha}{m_k}\vec{p}\,,\qquad
\vec{q}\,'=\vec{p}+\frac{M_\beta}{m_k}\vec{p}\,'\,,
\end{equation}
where $m_k$ is the mass of the exchanged particle (or quasiparticle).
After a partial wave decomposition, the equations (\ref{A160}) are reduced
to a set of integral equations in one variable, being thus manageable for
practical purposes.

Turning to the $\eta$-$3N$ system we will firstly define the
notation. The set of channels,
we are interested in, is given by (\ref{eq145}).
In what follows, we do not consider the explicit structure of the
equations, connected with the separable expansion of the basic
subamplitudes (index ``$n$'' in (\ref{A150}))
and drop also the spin-isospin indices. It is convenient to order the
nucleons always cyclically, 
since in this case we have not to trace the sign, when changing from one
state to another. Denoting the nucleons as particles 1, 2 and 3, we consider
the following 7 states which contribute
\begin{tabbing} xxx \=
xxxxxxxxxxxxxxxxxxxxxxxxxxxxx \=
xxxxxxxxxxxxxxxxxxxxxxxxxxxxxxxxxxxxxxxxxxxxxxxxxxxxxxxxxxx \kill
(i) \> one state in channel 1: \> $|\eta(123)\rangle$, \ denoted as 1\,, \\
 \> \> \\
(ii) \> three states in channel 2: \>
$|1(\eta 23)\rangle$,\ $|2(\eta 31)\rangle$,\ $|3(\eta 12)\rangle$,\
  denoted as $2_i$ with $i=1,2,3$ \,,\\
 \> \> \\
(iii) \> three states in channel 3: \>
$|(1\eta) (23)\rangle$,\ $|(2\eta) (31)\rangle$,\ $|(3\eta) (12)\rangle$,\
  denoted as $3_i$ with $i=1,2,3$\,.
\end{tabbing}

Here we assume that the wave functions of the subsystems containing two
and three nucleons are already antisymmetrized. For the present purpose we are
interested in those amplitudes, which describe the transitions from
channel 1 to the channels $1$, $2_i$ and $3_i$.
They will be denoted respectively as $X_1$, $X_{\alpha_i}$
with $i=1,2,3$. Then using the separable approximation for the
amplitudes (\ref{eq95}), (\ref{eq100}) and (\ref{eq140}),
the equations (\ref{A160}) take the form
\begin{eqnarray}\label{eqA10}
X_1&=&\sum\limits_{j=1}^3
Z_{1;2_j }\Theta_{2_j }X_{2_j }+ \sum\limits_{j=1}^3
Z_{1;3_j }\Theta_{3_j }X_{3_j }\,, \nonumber \\ %
X_{2_i }&=&Z_{2_i ;1}+Z_{2_i ;1}\Theta_1X_1+
\sum\limits_{j=1}^3 (1-\delta_{ij})Z_{2_i ;2_j }\Theta_{2_j }X_{2_j }+
\sum\limits_{j=1}^3 Z_{2_i ;3_j }\Theta_{3_j }X_{3_j }\,, \\
X_{3_i }&=&Z_{3_i ;1}+Z_{3_i ;1}\Theta_1X_1+
\sum\limits_{j=1}^3 Z_{3_i ;2_j }\Theta_{2_j }X_{2_j }\,,\nonumber
\end{eqnarray}
where $i=1,2,3$.
The meaning of the driving terms $Z_{\alpha_i\beta_j}$ is explained
schematically by the diagrammatical representation in
Fig.~\ref{fig2}. Their analytical expressions are given in Appendix B.
The terms $Z_{2_i ;3_j }$ and $Z_{3_i ;2_j }$ have different structures for
$i=j$ and for $i\ne j$ and therefore are written down separately.

The wave functions of different states belonging to the same channel may be
obtained from each other by cyclic permutation of the nucleon
coordinates. This fact allows one to obtain various relations between
the transition amplitudes $Z_{\alpha_i;\beta_j}$.
For example, we have by cyclic permutation
\begin{equation}\label{eqA15}
Z_{2_1 ;2_2 }=\langle 1(\eta 23)|Z|2(\eta 31)\rangle =
\langle 2(\eta 31)|Z|3(\eta 12)\rangle=Z_{2_2 ;2_3}\,.
\end{equation}
Repeating the procedure, one finds the general
relation for $i\neq j$
\begin{equation}\label{eqA15a}
Z_{2_i ;2_j }=Z_{2_1 ;2_2}\,.
\end{equation}
In the same manner and by applying a combination of a cyclic permutation
with a permutation within an antisymmetrized $NN$-state, one obtains for
the transition $2\to 3$ the general relations
\begin{eqnarray}
Z_{2_i;3_i}&=&Z_{2_1;3_1}\,,\\
Z_{2_i;3_j}&=&Z_{2_1;3_2}\quad \mbox{for }i\neq j\,.
\end{eqnarray}
Then, taking into account the obvious identities
\begin{equation}\label{eqA20}
\Theta_{\alpha_1 }=\Theta_{\alpha_2 }=\Theta_{\alpha_3 }
\equiv\Theta_\alpha\,, \ \ (\alpha=2,3)
\end{equation}
and defining the properly antisymmetrized amplitudes in the channels 1 to 3
by
\begin{equation}\label{eqA25}
X^a_1=X_1\,, \ \
X^a_{\alpha}=\frac{1}{\sqrt{3}}\sum\limits_{i=1}^3X_{\alpha_i }\,, \quad
\mbox{for } \alpha=2,3\,,
\end{equation}
we arrive at the following set of equations
\begin{eqnarray}\label{eqA30}
X^a_1&=&\sqrt{3} Z_{1;2_1 }\Theta_2X^a_2+
\sqrt{3}Z_{1;3_1 }\Theta_3 X^a_3\,,\nonumber \\
X^a_2&=&\sqrt{3}\, Z_{2_1 ;1}+\sqrt{3}\,Z_{2_1 ;1}\Theta_1X^a_1+
2Z_{2_1 ;2_2 }\Theta_2X^a_2 +
(2Z_{2_1 ;3_2 }+Z_{2_1 ;3_1 })\Theta_3 X^a_3\,, \\
X^a_3&=&\sqrt{3}\, Z_{3_1 ;1}+\sqrt{3}\,Z_{3_1 ;1}\Theta_1X^a_1+
(2Z_{3_1 ;2_2 }+Z_{3_1 ;2_1 })\Theta_2 X^a_2\,.\nonumber
\end{eqnarray}
The last step we have to make is to change from the driving terms
$Z_{\alpha_i;\beta_j}$ to the terms, which couple the antisymmetrized
states
\begin{equation}\label{eqA35}
Z_{12}=\frac{1}{\sqrt{3}}\langle \eta(123)|Z|\hat{P} 1(\eta 23)\rangle =
\langle \eta(123)|Z|1(\eta 23)\rangle=Z_{1;2_1 }\,.
\end{equation}
Here $\hat{P}$ stands for the cyclic permutations of the nucleon
labels according to
\begin{equation}
|\hat{P}i(\eta jk)\rangle = 
|i(\eta jk)\rangle +|j(\eta ki)\rangle +|k(\eta ij)\rangle \,.
\end{equation}
In a similar manner we obtain
\begin{eqnarray}\label{eqA40}
Z_{22}&=&\frac{1}{3}\langle \hat{P}1(\eta 23)|Z|\hat{P} 1(\eta 23)\rangle
= 2Z_{2_1 ;2_2 }\,,\nonumber
\\ Z_{21}&=&Z_{2_1 ;1}\,,\nonumber
\\
Z_{13}&=&\frac{1}{\sqrt{3}}\langle \eta(123)|Z|\hat{P} (1\eta)(23)\rangle
= Z_{1;3_1 }\,,\\
Z_{23}&=&\frac{1}{3}\langle \hat{P}1(\eta 23)|Z|\hat{P}
(1\eta)(23)\rangle = 2Z_{2_1 ;3_2 }+Z_{2_1 ;3_1 }\,,\nonumber \\
Z_{31}&=&Z_{3_1 ;1}\,,\nonumber \\
Z_{32}&=&2Z_{3_1 ;2_2 }+Z_{3_1 ;2_1 }\,.\nonumber
\end{eqnarray}
Substituting (\ref{eqA40}) into (\ref{eqA30}), we end up with the system
(\ref{eq150}).

\renewcommand{\theequation}{B\arabic{equation}}
\setcounter{equation}{0}
\section*{Appendix B}
\label{appB}

In this appendix we list the explicit expressions for the driving
terms, appearing in the separable-potential approach to the $\eta
$-$3N$ problem.  As was mentioned already, we consider only $s$-wave
orbitals in all
subsystems. Therefore the spin algebra may be done independently. The
corresponding spin-isospin coefficients $\delta_{ss'}$ and $\Lambda_{ss'}$
are found by applying the usual angular momentum recoupling schemes
(see e.g.~\cite{Edmon57}). All driving terms have the same general
functional form
\begin{equation}\label{eqB10}
\widetilde Z^{(ss')}_{\alpha\beta;nn'}(p,p',E)=\frac{\Omega_{ss'}}{2}\,
\int\limits_{-1}^{+1}
\widetilde u^{(s)}_n(q,{\cal E}-\frac{p^2}{2M_\alpha})\,
\widetilde \tau(E-f(p,p'\,))\,
\widetilde v^{(s)}_{n}
(q',{\cal E}'-\frac{p'^2}{2M_\beta})\,d(\hat{p}\cdot\hat{p}')\,.
\end{equation}
We list in Table~\ref{tab_app_B} for each driving term the
corresponding assignments for the various symbols appearing in
(\ref{eqB10}), where $\Lambda_{ss'}$ is defined by
\begin{eqnarray}
\Lambda&=&\left(
\begin{array}{rr}
\frac{1}{4} & -\frac{3}{4} \\
 & \\
-\frac{3}{4} & \frac{1}{4} \end{array}
\right)\,.
\end{eqnarray}
The coefficient 2 in the terms $Z_{22}$ and $Z^2_{23}$
stems from the identity of the nucleons, as described in Appendix A.
The reduced masses are defined in (\ref{eq165}),
(\ref{eq40}) and (\ref{eq65}).
Furthermore, one should note that we have split $Z_{23}$ into two terms
\begin{eqnarray}
Z_{23}=Z^1_{23}+Z^2_{23}\,.
\end{eqnarray}
The remaining driving terms are obtained from
\begin{eqnarray}\label{eqB75}
Z^{(ss')}_{21;nn'}(p,p',E)&=&Z^{(s's)}_{12;n'n}(p',p,E)\,,\\
Z^{(ss')}_{31;nn'}(p,p',E)&=&Z^{(s's)}_{13;n'n}(p',p,E)\,,  \\
Z^{(ss')}_{32;nn'}(p,p',E)&=&Z^{(s's)}_{23;n'n}(p',p,E)\,.
\end{eqnarray}

In accordance with the approximation (\ref{eq190}) for the target wave
function, we redefine the driving terms $Z_{21}$ and
$Z_{31}$ in (\ref{eq175}) as well as the terms $Z_{12}$ and $Z_{13}$ in
(\ref{eq185}). Instead of them, we introduce effective potentials
which are symmetrized over the singlet and triplet $NN$-states in the
triton
\begin{equation}\label{eqB80} Z^{(ss')}_{12;1n'}(p,p',E) \to
\overline{Z}^{(ss')}_{12;1n'}(p,p',E) = 
\frac{1}{2}\left(Z^{(00)}_{12;1n'}(p,p',E)-Z^{(11)}_{12;1n'}(p,p',E)\right)\,.
\end{equation}
The other terms
$\overline{Z}^{(ss')}_{21;n'1}$,
$\overline{Z}^{(ss')}_{13;1n'}$, and
$\overline{Z}^{(ss')}_{31;n'1}$
are defined by analogous relations.



\begin{table}
\renewcommand{\arraystretch}{2.0}
\caption{
Listing of parameters determining the separable parametrization of
the $NN$ and $\eta N$ scattering matrices.
}
\begin{center}
\begin{tabular}{ccccccccc}
$g_d^{(0)}$ [MeV$^{-1/2}$] & $g_d^{(1)}$ [MeV$^{-1/2}$] & $\beta_0$
[fm$^{-1}$] & $\beta_1$ [fm$^{-1}$] & $g_{N^*}^\eta$ & $g_{N^*}^\pi$ &
$\beta_\eta$ [fm$^{-1}$] & $\beta_\pi$ [fm$^{-1}$] & $M_0$ [MeV] \\ \hline
0.4076 & 0.4863 & 1.4488 & 1.4488 & 2.10 & 1.04 & 6.5 & 4.5 & 1661  \\
\end{tabular}
\label{tab1}
\end{center}
\end{table}

\begin{table}
\renewcommand{\arraystretch}{2.0}
\caption{
The scattering length $a_{\eta\, ^3\mathrm{H}}$ as a function of the
number $n_\alpha$ of separable terms kept in the Hilbert-Schmidt expansion
for the (1+3) and (2+2) amplitudes. The values of $n_1$, $n_2$ and
$n_3$ denote the number of terms kept in the separable expansion of
the amplitudes \protect{(\ref{eq95}), (\ref{eq100}) and
(\ref{eq140})}, respectively.} 
\begin{tabular}{cccc}
$n_1$ & $n_2$ & $n_3$ & $a_{\eta\, ^3\mathrm{H}}$ [fm] \\ \hline
2   &   2   &   2   &   $2.51+i\,1.60$ \\
4   &   2   &   2   &   $2.54+i\,1.66$ \\
4   &   4   &   2   &   $3.68+i\,3.46$ \\
4   &   4   &   4   &   $3.99+i\,4.74$ \\
6   &   4   &   4   &   $3.98+i\,4.75$ \\
6   &   6   &   4   &   $4.16+i\,5.52$ \\
6   &   6   &   6   &   $4.18+i\,5.67$ \\
6   &   8   &   6   &   $4.19+i\,5.69$ \\
\end{tabular}
\label{tab2}
\end{table}


\begin{table}
\renewcommand{\arraystretch}{2.0}
\caption{Comparison of the predictions for the $\eta\, ^3$H-scattering lengths
of the present model with the ones obtained in
\protect\cite{Wyce95,Belya95} for the corresponding 
$\eta N$-scattering length.}
\begin{tabular}{ccc}
$a_{\eta N}$ [fm] & $a_{\eta\, ^3\mathrm{H}}$ [fm]
& $a_{\eta\, ^3\mathrm{H}}$ [fm] (this work)   \\
\hline
$0.57+i\,0.39$   &  $1.32+i\,4.37$ \protect{\cite{Wyce95}}
& $2.23+i\,3.00$ \\
$0.29+i\,0.36$   &  $0.58+i\,2.17$ \protect{\cite{Wyce95}}
& $0.97+i\,1.72$ \\
$0.27+i\,0.22$   &  $0.41+i\,2.00$ \protect{\cite{Belya95}}
& $0.69+i\,0.67$ \\
$0.55+i\,0.30$   & $\!\!\!\!\!-1.56+i\,3.00$ \protect{\cite{Belya95}}
& $2.35+i\,1.68$ 
\end{tabular}
\label{tab3}
\end{table}


\begin{table}
\renewcommand{\arraystretch}{2.0}
\caption{
Listing of symbols appearing in (\ref{eqB10}).
}
\begin{center}
\begin{tabular}{cccccccc}
$\widetilde Z^{(ss')}_{\alpha\beta;nn'}$ & $\Omega_{ss'}$ &
$\widetilde u^{(s)}_n$ &
$\vec{q}$ & $\widetilde \tau$ & $f(p,p'\,)$ &
$\widetilde v^{(s)}_{n}$ &
$\vec{q}\,'$ \\
\hline
$Z^{(ss')}_{12;nn'}$ & $\delta_{ss'}$ & $u^{(s)}_{n}$ & $\vec{p}\,'+\frac{1}{3}\vec{p}$ &
$\tau_d^{(s)}$ & $\frac{p^2}{2\tilde\mu_d}+\frac{3p'^2}{4M_N}
+\frac{\vec{p}\cdot\vec{p}\,'}{2M_N}$ & $v^{(s)}_{d;n}$ &
$\vec{p}+\frac{\tilde\mu_d}{2M_N}\vec{p}\,'$ \\
$Z^{(ss')}_{13;nn'}$ & $\delta_{ss'}$ & $u^{(s)}_{n}$ & $\vec{p}\,'+\frac{2}{3}\vec{p}$ &
$\tau_d^{(s)}$ & $\frac{p^2}{2\mu_{\eta N}}+\frac{3p'^2}{4M_N}
+\frac{\vec{p}\cdot\vec{p}\,'}{M_N}$ & $w^{(s)}_{d;n}$ &
$\vec{p}+\frac{\mu_{\eta N}}{M_N}\vec{p}\,'$\\
$Z^{(ss')}_{22;nn'}$ & $2\Lambda_{ss'}$ & $v^{(s)}_{N^*;n}$ &
$\vec{p}\,'+\frac{\tilde\mu_{N^*}}{M_N+m_\eta}\vec{p}$ &
$\tau_{N^*}$ &
$\frac{p^2}{2\tilde\mu_{N^*}}+\frac{p'^2}{2\tilde\mu_{N^*}}+\frac{\vec{p}\cdot\vec{p}\,'}{M_N+m_\eta}$
& $v^{(s')}_{N^*;n}$ &
$\vec{p}+\frac{\tilde\mu_{N^*}}{M_N+m_\eta}\vec{p}$\\
$Z^{1(ss')}_{23;nn'}$ & $\delta_{ss'}$ & $v_{d;n}^{(s)}$ & $\vec{p}\,'+\frac{\tilde\mu_d}{m_\eta}\vec{p}$ & $\tau_d^{(s)}$ & $\frac{p^2}{2\mu_{\eta N}}+\frac{p'^2}{2\tilde\mu_d}
+\frac{\vec{p}\cdot\vec{p}\,'}{m_\eta}$ & $w^{(s)}_{d;n}$ & $\vec{p}+\frac{\mu_{\eta N}}{m_\eta}\vec{p}\,'$\\
$Z^{2(ss')}_{23;nn'}$ & $2\Lambda_{ss'}$ & $v^{(s)}_{N^*;n}$ & $\vec{p}\,'+\frac{\tilde\mu_{N^*}}{M_N}\vec{p}$ & $\tau_{N^*}$ & $\frac{p^2}{M_N}+\frac{p'^2}{2\tilde\mu_{N^*}}
+\frac{\vec{p}\cdot\vec{p}\,'}{M_N}$ & $w^{(s')}_{N^*;n}$ & $\vec{p}+\frac{1}{2}\vec{p}\,'$\\
\end{tabular}
\label{tab_app_B}
\end{center}
\end{table}


\begin{thebibliography}{99}

\bibitem{Peng89}
J.C.\ Peng {\it et al.}, Phys. Rev. Lett. {\bf 63}, 2353 (1989).

\bibitem{May96}
B.\ Mayer {\it et al.}, Phys. Rev.\ C {\bf 53}, 2068 (1996).

\bibitem{Kru02}
B.\ Krusche, Contribution presented at Meson 2002, 7th
International Workschop on Production, Properties and Interaction of
Mesons (Cracow, Poland, May 24-28, 2002).

\bibitem{Wilkin93}
C.\ Wilkin, Phys. Rev.\ C {\bf 47}, R938 (1993).

\bibitem{Wyce95}
S.\ Wycech, A.M.\ Green, and 
J.A.\ Niskanen, Phys. Rev.\ C {\bf 52}, 544 (1995).

\bibitem{Belya95}
V.B.\ Belyaev, S.A.\ Rakityansky, S.A.\ Sofianos, M.\ Braun, and W.\ Sandhas,
Few Body Syst., Suppl.\ {\bf 8}, 309 (1995).
S.A.\ Rakityansky, S.A.\ Sofianos, M.\ Braun, V.B.\ Belyaev, and W.\ Sandhas,
Phys. Rev.\ C {\bf 53}, R2043 (1996).

\bibitem{Shev01}
N.V.\ Shevchenko, V.B.\ Belyaev, S.A.\ Rakityansky, S.A.\ Sofianos, and
W.\ Sandhas, nucl-th/0108031.

\bibitem{Khem02}
K.P.\ Khemchandani, N.G.\ Kelkar, and B.K.\ Jain, nucl-th/0112065.

\bibitem{Yak67}
O.A.\ Yakubovsky, Yad. Fiz. {\bf 5}, 1312 (1967)
(Trans.: Sov. J. Nucl. Phys.\ {\bf 5}, 937 (1967)).

\bibitem{Fad70}
L.D.\ Faddeev, Proc. First Int. Conf. on the Three-Body Problem, Birmingham, 
1969, eds. J.S.C. McKee and P.M. Rolph (North-Holland, Amsterdam 1970).

\bibitem{GS67}
P.\ Grassberger and W.\ Sandhas, Nucl. Phys.\ {\bf B2}, 181 (1967).

\bibitem{AGS70}
E.O.\ Alt, P.\ Grassberger, and W.\ Sandhas, Phys. Rev.\ C {\bf 1}, 85 (1970).

\bibitem{Fons86}
A.C.\ Fonseca, Proc.\ 8th Autumn School on Models and Methods in
Few-Body Physics, Lisboa, 1986, eds. L.S.\ Ferreira, A.C.\ Fonseca
and L.\ Streit, Lecture Notes in Physics {\bf 273}, 161 (1986). 

\bibitem{Narod81}
I.M.\ Narodetsky, Riv. Nuovo Cim.\ {\bf 4}, 1 (1981).

\bibitem{Tjon75}
J.A.\ Tjon, Phys. Lett.\ {\bf B56}, 217 (1975); 
Phys. Lett.\ {\bf B63}, 391 (1976).

\bibitem{Fons99}
A.C.\ Fonseca, Phys. Rev. Lett.\ {\bf 83}, 4021 (1999). 

\bibitem{Carb99}
F.\ Ciesielski, J.\ Carbonell, and C.\ Gignoux, Phys. Lett.\ {\bf B447},
199 (1999).  
\bibitem{Viv01}
M.\ Viviany, A.\ Kievsky, S.\ Rosati, E.A.\ George, and L.D.\ Knutson,
Phys. Rev. Lett.\ {\bf 86}, 3739 (2001). 

\bibitem{Avish83}
Y.\ Avishai and T.\ Mizutani, Nucl. Phys.\ {\bf A393}, 429 (1983).

\bibitem{Ueda89}
T.\ Ueda, Nucl. Phys.\ {\bf A505}, 610 (1989).

\bibitem{Cant98}
L.\ Canton, Phys. Rev.\ C {\bf 58}, 3121 (1998).

\bibitem{Yam54}
Y.\ Yamaguchi, Phys. Rev.\ C {\bf 95}, 1628 (1954).

\bibitem{Batin95}
M.\ Batinic, I.\ Slaus, A.\ Svarc, and 
B.\ Nefkens, Phys. Rev.\ C {\bf 55}, 2310
(1995).

\bibitem{Wyce97}
A.M.\ Green and S.\ Wycech, Phys. Rev.\ C {\bf 55}, R2167 (1997).

\bibitem{Wein64}
S.\ Weinberg, Phys. Rev.\ {\bf 133}, B232 (1964).

\bibitem{Sitenko}
A.G.\ Sitenko, {\it Lectures in Scattering Theory}
(Pergamon Press, Oxford, 1974)

\bibitem{Love64}
C.\ Lovelace, Phys. Rev.\ {\bf 135}, B1225 (1964).

\bibitem{Shev98}
N.V.\ Shevchenko, S.A.\ Rakityansky, S.A.\ Sofianos, V.B.\ Belyaev, and
W.\ Sandhas, Phys. Rev.\ C {\bf 58}, R3055 (1998).

\bibitem{FiAr01}
A.\ Fix and H.\ Arenh\"ovel, Nucl. Phys. {\bf A697}, 277 (2001).

\bibitem{Wyce01}
A.M.\ Green and S.\ Wycech, Phys. Rev.\ C {\bf 64}, 045206 (2001).

\bibitem{Schiff64}
L.I.\ Schiff, Phys. Rev.\ {\bf 133}, B802 (1964).

\bibitem{Tryas97}
V.A.\ Tryasuchev, Yad. Fiz. {\bf 60}, 245 (1997)
(Trans.: Phys. Atom. Nucl.\ {\bf 60}, 186 (1997)).

\bibitem{Edmon57}
A.R.\ Edmonds, {\it Angular Momentum in Quantum Mechanics}
(Princeton University Press, Princeton, New Jersey, 1957).

\end{thebibliography}
\end{document}